\documentclass[journal,twoside,web]{ieeecolor}

\usepackage{main}
\usepackage{cite}

\usepackage{amsmath,amssymb,amsfonts}
\usepackage{algorithmic}
\usepackage{graphicx}
\usepackage{textcomp}

\usepackage{multicol}
\usepackage{multirow}
\usepackage{booktabs}
\usepackage[ruled]{algorithm2e}
\usepackage{booktabs}
\usepackage{utfsym}
\usepackage{bm}
\usepackage{pifont}
\usepackage{setspace}

\def\BibTeX{{\rm B\kern-.05em{\sc i\kern-.025em b}\kern-.08em
    T\kern-.1667em\lower.7ex\hbox{E}\kern-.125emX}}

\begin{document}
\title{Model-based Federated Learning for Accurate MR Image Reconstruction from Undersampled k-space Data}

\author{Ruoyou Wu, Cheng Li, Juan Zou, Qiegen Liu, Hairong Zheng and Shanshan Wang
\thanks{This research was partly supported by the National Natural Science Foundation of China (62222118, U22A2040), Guangdong Provincial Key Laboratory of Artificial Intelligence in Medical Image Analysis and Application (2022B1212010011), and Shenzhen Science and Technology Program (RCYX20210706092104034, JCYJ20220531100213029)}
\thanks{R. Wu is with the Paul C. Lauterbur Research Center for Biomedical Imaging, Shenzhen Institute of Advanced Technology, Chinese Academy of Sciences, Shenzhen 518055, China, and with the Peng Cheng Laboratory, Shenzhen 518055, China, and with the University of Chinese Academy of Sciences, Beijing 100049, China and also with Guangdong Provincial Key Laboratory of Artificial Intelligence in Medical Image Analysis and Application, China. E-mail: ry.wu@siat.ac.cn.}
\thanks{C. Li and H. Zheng are with the Pual C. Lauterbur Research Center for Biomedical Imaging, Shenzhen Institute of Advanced Technology, Chinese Academy of Sciences, Shenzhen 518055, China. E-mail: ({cheng.li6, hr.zheng}@siat.ac.cn).}
\thanks{J. Zou is with the School of Physics and Optoelectronics, Xiangtan University, Xiangtan 411105, China and 
 also with the Pual C. Lauterbur Research Center for Biomedical Imaging, Shenzhen Institute of Advanced Technology, Chinese Academy of Sciences, Shenzhen 518055, China. E-mail: zjuan@xtu.edu.cn.}
 \thanks{Q. Liu is with the Department of Electronic Information Engineering, Nanchang University, Nanchang 330031, China. E-mail: liuqiegen@ncu.edu.cn.}
 \thanks{S. Wang is with the Paul C. Lauterbur Research Center for Biomedical Imaging, Shenzhen Institute of Advanced Technology, Chinese Academy of Sciences, Shenzhen 518055, China. E-mail: sophiasswang@hotmail.com (Corresponding author: Shanshan Wang)}
 }

\maketitle

\begin{abstract}
Deep learning-based methods have achieved encouraging performances in the field of magnetic resonance (MR) image reconstruction. Nevertheless, to properly learn a powerful and robust model, these methods generally require large quantities of data, the collection of which from multiple centers may cause ethical and data privacy violation issues. Lately, federated learning has served as a promising solution to exploit multi-center data while getting rid of the data transfer between institutions. However, high heterogeneity exists in the data from different centers, and existing federated learning methods tend to use average aggregation methods to combine the client's information, which limits the performance and generalization capability of the trained models. In this paper, we propose a Model-based Federated learning framework (ModFed). ModFed has three major contributions: 1) Different from the existing data-driven federated learning methods, model-driven neural networks are designed to relieve each client’s dependency on large data; 2) An adaptive dynamic aggregation scheme is proposed to address the data heterogeneity issue and improve the generalization capability and robustness the trained neural network models; 3) A spatial Laplacian attention mechanism and a personalized client-side loss regularization are introduced to capture the detailed information for accurate image reconstruction. ModFed is evaluated on three in-vivo datasets. Experimental results show that ModFed has strong capability in improving image reconstruction quality and enforcing model generalization capability when compared to the other five state-of-the-art federated learning approaches. Codes will be made available at https://github.com/ternencewu123/ModFed.
\end{abstract}


\section{Introduction}
\label{sec:introduction}
\IEEEPARstart{M}{AGNETIC} Resonance Imaging (MRI) is widely used in radiology and medicine as a non-invasive imaging modality. One prominent problem of MRI is its long scanning time when compared to other imaging techniques. To speed up MRI data acquisition\cite{fessler2020optimization}, high-quality MR image reconstruction from undersampled k-space data has been extensively investigated. 
Among them, deep learning has been a very fascinating direction to facilitate this task by drawing valuable knowledge from large datasets\cite{hammernik2018learning,wang2016accelerating,zhu2018image,schlemper2017deep,aggarwal2018modl}. For example, there are image domain based deep learning methods which try to learn prior knowledges or the direct mapping  between undersampled and fully sampled data in the image domain, showing strong capability in removing the aliasing noise and artifacts for the undersampled MR image reconstruction\cite{wang2020deepcomplexmri,lee2018deep,han2018deep,mardani2018deep,quan2018compressed,feng2021donet,qin2018convolutional}. There are also k-space domain learning methods that attempt to full-fill k-space measurements using deep neural networks, which could capture detailed information for the final reconstruction\cite{akccakaya2019scan,han2019k,sriram2020grappanet}. Also, there are works trying to accomplish MR image reconstruction by learning in hybrid domains (k-space and image domain), which simultaneously explore the advantages of both domains \cite{eo2018kiki,wang2022dimension}. However, deep learning-based approaches typically require large amounts of high-quality training data for training robust models. 

To exploit big data without violating patient's privacy, federated learning has become a very popular tool \cite{kaissis2020secure}. It provides a platform for different institutions to collaboratively learn models by using local computing resources and data without releasing local private data to the public \cite{guo2021multi,feng2022specificity,elmas2022federated,levac2023federated,feng2023learning,dalmaz2022one}. Federated learning generally consists of the server and client components. Specifically, the server and client communicate periodically to aggregate information and get a global model. The global model is then shared with all clients. Communication in federated learning only involves model parameters or gradient information without aggregating data from different institutions, which facilitates the multi-institutional collaboration and alleviates the privacy concerns \cite{li2020federated}. Specifically, FedAvg \cite{mcmahan2017communication} is the most widely used federated learning algorithm, which aggregates the weights of multiple institutions by averaging them and then dispatches them to different institutions for the next communication. Building on FedAvg, there are some studies trying to develop different strategies to obtain a better global model. For example, Li et al. \cite{li2020fedprox} obtained better performance by re-parameterization of FedAvg. Li et al. \cite{li2021fedbn} introduced a FedBN method, which can alleviate the feature shift by local batch normalization. In MR image reconstruction, Guo et al. \cite{guo2021multi} proposed a multi-institutional collaborative MRI reconstruction method to improve the similarity of potential spatial representations through adversarial alignment between source and target sites, but the communications cost of this approach is relatively high.  Furthermore, in practice, MR data acquired by different institutions can be highly heterogeneous because of employing different scanning devices, sampling modalities, or acceleration rates \cite{knoll2020fastmri}. The data heterogeneity increases the difficulty of adapting the global model to local data, which affects the model convergence and leads to degraded model performances \cite{karimireddy2020scaffold}. Although these methods partially alleviate the problem of data heterogeneity, most of them aim to learn a global model and seldom consider the personalized characteristics of the client, which may reduce the performance of the client.

Personalized federated learning (PEL) frameworks have been particularly proposed to address the data heterogeneity problem \cite{zhao2022personalized,feng2022specificity}. In contrast to classical federated learning, which pursues a global model, personalized federated learning aims to simultaneously learn a customized model for each client, so that the learned local model can be better adapted to the client’s data distribution. Liang et al. \cite{liang2020think} proposed LG-FedAvg, which firstly learns generic representations and then each client trains a specific local-head on local private data. In contrast, Arivazhagan et al. \cite{arivazhagan2019federated} constructed a federated learning framework with base layer and personalization layer, where the base layer is used to learn generic representations and the personalization layer is used to learn the intrinsic representations of local private data. Moreover, Zhao et al. \cite{zhao2022personalized} proposed a few-shot federated learning method that can achieve client personalization with only a small amount of data. In MR image reconstruction, Feng et al. \cite{feng2022specificity} proposed a FedMRI method, which separates the reconstruction model into a client-shared global encoder and a client-specific decoder to obtain good performances in local client reconstruction. However, the server still uses the average aggregation method, which limits the performance of the model to some extent. Furthermore, to the best of our knowledge, the existing federated learning MR image reconstruction frameworks are purely data-driven approaches, which lack physical knowledge's guidance and still rely on big data on the client side. Last but not least, these methods are very sensitive to noise due to the pure reliance on the dataset.

To address the above mentioned issues, this paper proposes a model-based federated learning framework (ModFed), which has the following three main contributions:

1) Unlike the existing pure data-driven federated learning MR image reconstruction methods, we utilize model-driven neural networks that exploit the MR physics knowledge and reduce the client’s dependence on the large amount of data.  

2) Instead of utilizing the average operation, we propose a dynamic adaptive aggregation scheme to address the data heterogeneity issue and improve the generalization capability and robustness of the trained model.

3) A spatial Laplacian attention mechanism and a personalized client-side loss regularization are introduced to capture data distribution, fine structure and details for accurate MR image reconstruction from undersampled data.

The rest of this article is organized as follows. Section 2 recaps the MR imaging and federated learning-based reconstruction basics. Section 3 describes the details of our proposed method, ModFed. Section 4 reports and discusses the experimental results. Section 5 concludes this article.

\section{Preliminary}
\subsection{MR Imaging Basics}
The acquisition process of MRI data is often represented as:
\begin{equation}
\mathbf{b=Am}+\bm{\epsilon}
\end{equation}
where $\mathbf{b}\in \mathbb{C}^{M}$ indicates the measurement data;  $\mathbf{m}\in \mathbb{C}^{N}$ represents the image to be reconstructed and $\bm{\epsilon} \in \mathbb{C}^{M}$ represents measurement noise; $\mathbf{A=PF}$ represents the forward operator with $\mathbf{P}$ and $\mathbf{F}$  respectively denoting the undersampled matrix and the Fourier transform. The main purpose of MR image reconstruction is to efficiently learn the mapping relationship between the undersampled data and the fully sampled data \cite{wang2021deep}. To obtain the image from the measurements, we normally tend to solve the following   inverse problem:
\begin{equation}
    \mathbf{\hat{m}}=arg\min_{\mathbf{m}}\frac{1}{2}\left \| \mathbf{b-Am} \right \|_{2}^{2}+\lambda R(\mathbf{m})
\end{equation}
where $R(\mathbf{m})$ represents the introduced regularization term to conquer the ill-posed nature. $\lambda$ denotes the corresponding penalty weight. In this paper, we use the iterative method in literature \cite{aggarwal2018modl} to solve Eq. (2). The specific process is described as follows:
\begin{equation}
\left\{
    \begin{array}{lr}
    \mathbf{r}^{j}=E_{\bm{\omega}}\left (\mathbf{m}^{j}  \right ) &  \\
    \mathbf{m}^{j+1}=\left (\mathbf{A}^{H}\mathbf{A}+\lambda \mathbf{I}  \right ) ^{-1}\left (\mathbf{A}^{H}\mathbf{b} + \lambda \mathbf{r}^{j}\right )
    \end{array}
\right.
\end{equation}
where $E_{\bm{\omega}}(\cdot)$ denotes a learnable network, $\mathbf{r}^{j}$ denotes an intermediate iterative variable, and $\mathbf{A}^{H}$ represents the conjugate operator of $\mathbf{A}$. The data consistency term is solved using the conjugate gradient algorithm, and $\lambda$ can also be adaptively learned using the network.

\subsection{FL-based MR Image reconstruction}
In the centralized setting, by using paired datasets $\left \{ (\mathbf{b}_{i},\mathbf{m}_{i})\mid i=1,..., N \right \}$, Eq. (3) can be optimized by minimizing the following loss:
\begin{equation}
    \mathcal{L}_{rec}=\frac{1}{N}\sum_{i=1}^{N}\left \| f(\mathbf{A}^{H}\mathbf{b}_{i};\Theta)-\mathbf{m}_{i} \right \|_{2} 
\end{equation}
where $\Theta$ represents the learnable network weights, $\mathbf{m}_{i}$ denotes the reference data of the $i^{th}$ sample, $N$ represents the number of training samples.

When the client only has limited training data, it becomes difficult to train a model with satisfactory performance. Federated learning obtains a better model while protecting client data privacy through communication between multiple clients and a central server \cite{li2020federated}. In federated learning, it is assumed that there are $K$ hospitals/clients. Let $\mathcal{D}^{1},\mathcal{D}^{2},\cdots,\mathcal{D}^{K}$ denote the local datasets of different clients. In each communication round, the central server broadcasts the global model to the local model, $\Theta_{C}\to\Theta_{C^{k}}$, and then each client trains the local model by minimizing the following loss:
\begin{equation}
    \mathcal{L}_{rec}^{k}=\mathbb{E}_{(\mathbf{b}^{k},\mathbf{m}^{k})\sim \mathcal{D}^{k}}\left [ \left \| f_{k}(\mathbf{A}^{H}\mathbf{b}^{k};\Theta_{C^{k}})-\mathbf{m}^{k} \right \| _{2} \right ] 
\end{equation}
where $\mathcal{L}_{rec}^{k}$ denotes the training loss of the $k^{th}$ client and $f_{k}(\cdot)$ represents the $k^{th}$ local model parameterized by $\Theta_{C^{k}}$. $(\mathbf{b}^{k},\mathbf{m}^{k})$ denotes the training data pair of the $k^{th}$ client. Each client is optimized by stochastic gradient descent in local training:

\begin{equation}
    \Theta_{C^{k}}^{t,z+1}=\Theta_{C^{k}}^{t,z}-\eta_{k}\bigtriangledown\mathcal{L}_{rec}^{k}(\mathbf{A}^{H}\mathbf{b}^{k};\Theta_{C^{k}}^{t,z}) 
\end{equation}
where $t$ denotes the global epoch; $z$ denotes the local epoch and $\eta_{k}$ represents the learning rate of the $k^{th}$ client. Then, all clients send the trained weights to the central server. In federated learning-based MR image reconstruction, the global model weight is often obtained by aggregating local weights through the federated averaging (FedAvg) \cite{mcmahan2017communication} method:

\begin{equation}
    \Theta_{C}^{t}=\sum_{k=1}^{K}\frac{N_{k}}{N}\Theta_{C^{k}}^{t}  
\end{equation}
where $\Theta_{C}^{t}$ denotes the server weights in the $t^{th}$ global communication round; $N_{k}$ denotes the amounts of training data for the  $k^{th}$ client and $\Theta_{C^{k}}^{t}$ represents the weights of the $k^{th}$ client at the $t^{th}$ global communication round. The client and server communicate $T$ times to get the final global model $\Theta_{C}^{T}$. The trained global model $f_{\Theta_{C}}(\cdot)$ is then used in the testing process:

\begin{equation}
    \hat{\mathbf{m}}_{i}^{k}=f_{\Theta_{C}}(\mathbf{A}^{H}\mathbf{b}_{i}^{k}) 
\end{equation}
where $\hat{\mathbf{m}}_{i}^{k}$ denotes the $i^{th}$ reconstructed image of the $k^{th}$ client, $\mathbf{b}_{i}^{k}$ is the undersampled k-space data of the $i^{th}$ sample at the $k^{th}$ client.

\section{Proposed Method}
\subsection{Architecture of ModFed}
Compared to the traditional data-driven federated learning MR image reconstruction approach, we propose a model-driven federated learning framework, called ModFed, using the model-driven neural networks, which reduces the dependence on large data. The overall framework is illustrated in Fig. 1. The convolutional kernels of different layers in a convolutional neural network extract different levels of information. The shallow layers mainly extract generic features, while the deep layers mainly extract semantic features \cite{zeiler2014visualizing}. Personalized federated learning is realized by global-shared parameters and local-personalized parameters. In addition, we have three major contributions: 1) Model-driven neural networks are designed to relieve each client’s dependency on large data by introducing physics priors; 2) An adaptive dynamic aggregation scheme is proposed to address the data heterogeneity and improve the generalization capability and robustness of the global model; 3) A spatial Laplacian attention mechanism and a personalized client-side loss regularization are introduced to capture the detailed information for accurate image reconstruction.

\begin{figure*}[htbp]
    \centering
    \setlength{\abovecaptionskip}{0.cm}
    \includegraphics[width=14cm, keepaspectratio]{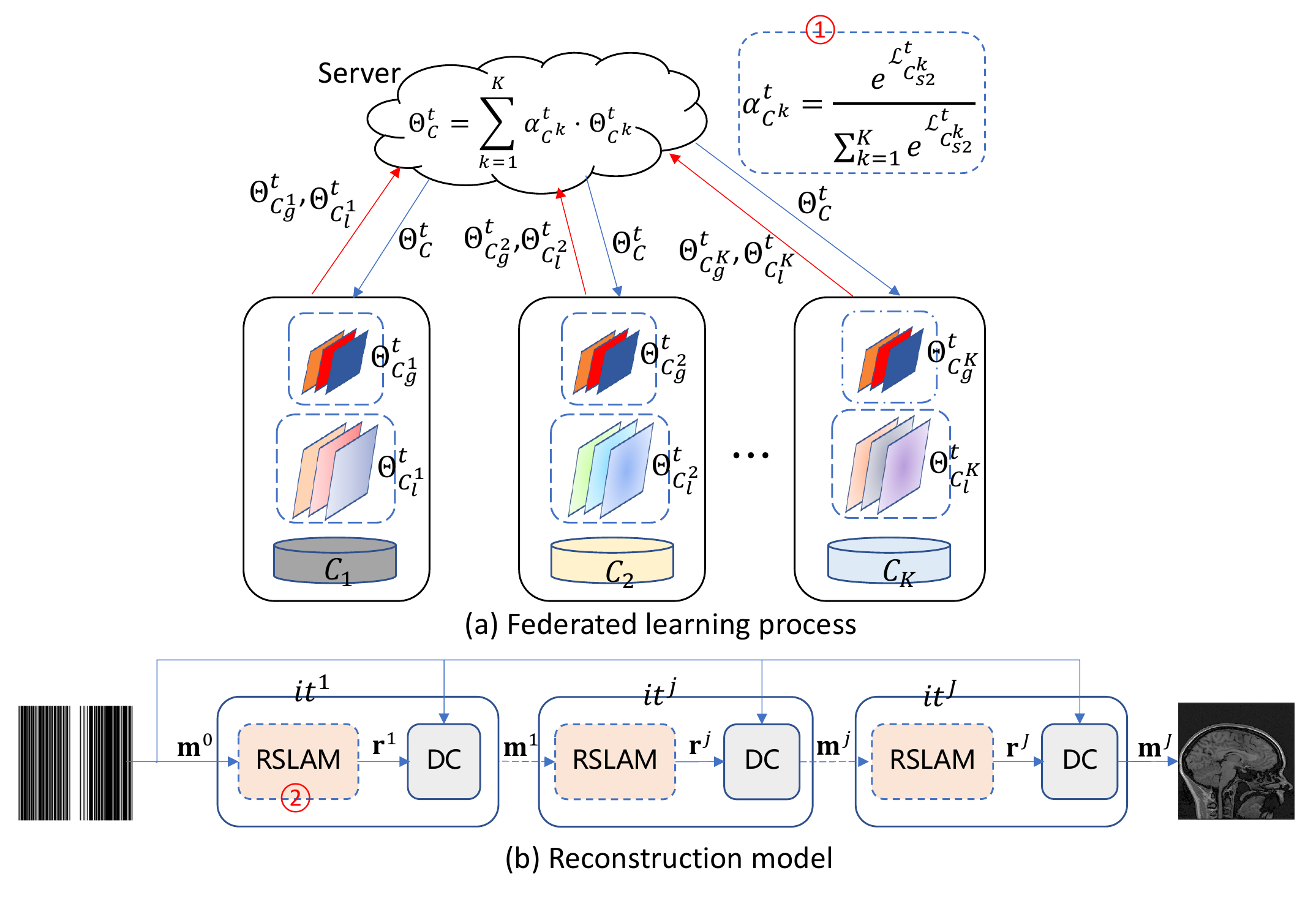}
    \caption{
    An overview of the ModFed framework. (a) Federated learning process. Instead of using the average aggregation method, adaptive dynamic aggregation scheme is used as shown in {\normalsize{\color{red}\ding{172}}}. Client parameters include global-shared parameters $\Theta_{C_{g}^{k}}^{t}$ and local-personalized parameters $\Theta_{C_{l}^{k}}^{t}$. (b) Reconstruction model. {\normalsize{\color{red}\ding{173}}} consists mainly of a spatial Laplacian attention module for capturing the detailed information of the image, which will be described in Section 3.2.
    }
    \label{fig1}
\end{figure*}

The ModFed model aims to learn the locally optimal personalization information for each client, while the server mainly learns the overall distribution of data. For this purpose, each communication round of ModFed alternates between client training and server updating. We divide each client model into two parts, $\Theta_{C_{g}^{k}}$ and $\Theta_{C_{l}^{k}}$, for global-shared and local-personalized learning, respectively. To this end, Eq. (4) can be rewritten as:

\begin{equation}
    \begin{split}
    \mathcal{L}_{C_{s1}^{k}} =  \mathbb{E}_{(\mathbf{b}^{k},\mathbf{m}^{k})\sim \mathcal{D}^{k}}[  \| f_{k}(\mathbf{A}^{H}\mathbf{b}^{k};\Theta_{C_{g}^{k}},\Theta_{C_{l}^{k}}) 
    \\
     - \mathbf{m}^{k} \|_{2} ], (\mathbf{b}^{k}, \mathbf{m}^{k}) \in s1
    \end{split}      
\end{equation}
where $\Theta_{C_{g}^{k}}$ and $\Theta_{C_{l}^{k}}$ denote the global-shared parameters and local-personalized parameters of client $k$, respectively. $s1$ denotes training subset 1. The updated rules of the server and clients are as follows:

a) \textbf{Client Updating Phase:} The server sends the global parameters $\Theta_{C}$ to each client, which is trained using the local private dataset and updates the local parameters by the stochastic gradient descent algorithm. To achieve enhanced reconstruction performance on each client, we introduce a personalized client-side loss regularization for each client, the loss of the client $k$ is:

\begin{equation}
    \mathcal{L}_{C_{rec}^{k}}=\mathcal{L}_{C_{s1}^{k}}+\gamma_{k}\mathcal{L}_{C}^{k}
\end{equation}
where $\gamma_{k}$ denotes the weighting factor to adjust the weights of $\mathcal{L}_{C}^{k}$; $\mathcal{L}_{C_{rec}^{k}}$ denotes the loss of client $k$, which consists of two parts: one is the loss $\mathcal{L}_{C_{s1}^{k}}$ of client $k$ on the training subset 1 and the other is the loss $\mathcal{L}_{C}^{k}$ of the server on training subset 2 of client $k$, which is mainly used to enhance reconstruction performance on each client. The calculation process is as follows: 

\begin{equation}
    \begin{split}
    \mathcal{L}_{C}^{k} =  \mathbb{E}_{(\mathbf{b}^{k},\mathbf{m}^{k})\sim \mathcal{D}^{k}}[  \| f_{C}(\mathbf{A}^{H}\mathbf{b}^{k};\Theta_{C}) 
    \\
     - \mathbf{m}^{k} \|_{2} ], (\mathbf{b}^{k}, \mathbf{m}^{k}) \in s2
     \end{split}
\end{equation}
where $\Theta_{C}=\Theta_{C_{g}}\cup \Theta_{C_{l}}$, $s2$ denotes training subset 2, and the client’s parameters update process is:

\begin{equation}
    \Theta_{C^{k}}^{t,z+1}= \Theta_{C^{k}}^{t,z}-\eta_{k}\bigtriangledown\mathcal{L}_{C_{rec}^{k}} 
\end{equation}
where $\Theta_{C^{k}}^{t,z}= \Theta_{C_{g}^{k}}^{t,z}\cup \Theta_{C_{l}^{k}}^{t,z}$, to realize local-personalized learning of $\Theta_{C_{l}^{k}}^{t,z}$ while ensuring the global sharing of $\Theta_{C_{g}^{k}}^{t,z}$.

b) \textbf{Server Aggregation Phase:} In each communication round, all clients send the trained weights $\Theta_{C^{k}}^{t}$ and the loss $ \mathcal{L}_{C_{s2}^{k}}$ (Eq. 15) of client $k$ on the training subset 2 to the server, which is updated as follows:

\begin{equation}
    \Theta_{C}^{t}= \sum_{k=1}^{K}\alpha_{C^{k}}^{t}\cdot \Theta _{C^{k}}^{t} 
\end{equation}
where $\Theta_{C}^{t}= \Theta _{C_{g}}^{t} \cup \Theta _{C_{l}}^{t}$, $\Theta _{C^{k}}^{t}=\Theta _{C_{g}^{k}}^{t}\cup \Theta _{C_{l}^{k}}^{t}$; $\alpha_{C^{k}}^{t}$ represents the aggregation weight of each client $k$, which is initialized by the data percentage of each client at the beginning of model training.

c) \textbf{Adaptive Dynamic Aggregation Scheme:} The federated average aggregation approach is suitable for client with similar data distribution, the solution of the federated average aggregation approach will deviate from the optimal solution when the client has a data heterogeneity problem, leading to a significant decline in the performance of the global model \cite{karimireddy2020scaffold}. Compensating the data heterogeneity of each client to improve the generalization capability and robustness of the global model, we introduce adaptive dynamic aggregation scheme, which is calculated as follows:

\begin{equation}
    \alpha _{C^{k}}^{t}=\frac{e^{\mathcal{L}_{C_{s2}^{k}}^{t} }}{\sum_{k=1}^{K}e^{\mathcal{L}_{C_{s2}^{k}}^{t} } } 
\end{equation}
where $ \alpha_{C^{k}}^{t}$ denotes the aggregation weight of client $k$ in the $t^{th}$ communication round and satisfies $\sum_{k=1}^{K}\alpha _{C^{k}}^{t}=1,0< \alpha _{C^{k}}^{t}<1$. $\mathcal{L}_{C_{s2}^{k}}^{t}$ denotes the loss of client $k$ on the training subset 2 in the $t^{th}$ communication, which can be calculated as follows:

\begin{equation}
    \begin{split}
    \mathcal{L}_{C_{s2}^{k}} =  \mathbb{E}_{(\mathbf{b}^{k},\mathbf{m}^{k})\sim \mathcal{D}^{k}}[  \| f_{k}(\mathbf{A}^{H}\mathbf{b}^{k};\Theta_{C_{g}},\Theta_{C_{l}^{k}}) 
    \\
     - \mathbf{m}^{k} \|_{2} ], (\mathbf{b}^{k}, \mathbf{m}^{k}) \in s2
    \end{split}      
\end{equation}
 where the loss $ \mathcal{L}_{C_{s2}^{k}}$ of client $k$ is larger,  it may mean that the data distribution on this client is far away from the data distribution on other clients. Therefore, we give it a larger aggregation weight to correct the data distribution. 
 
 The specific training process is shown in \textbf{Algorithm 1}. In global training, the server sends the aggregated weights $\Theta_{C}$ (Eq. 13) to each client. $\Theta_{C}$ is used by the client to calculate the loss $\mathcal{L}_{C}^{k}$ on the training subset 2 and as a part of the training loss, $\Theta_{C_{g}}\cup \Theta_{C_{l}^{k}}$ is used to calculate the loss $\mathcal{L}_{C_{s2}^{k}} $ on the training subset 2 for the aggregation weight. Then, each client updates the local gradient using the local dataset (Eq. 12). Finally, each client sends the training weights $\Theta_{C_{k}}$ and the loss $\mathcal{L}_{C_{s2}^{k}}$ to the server, which compute the aggregation weight (Eq. 14) based on $\mathcal{L}_{C_{s2}^{k}}$ and then calculate $\Theta_{C}$ for the next training according to Eq. (13). 

\begin{algorithm}[htbp]
 \caption{ModFed}\label{morel}
  \SetKwInOut{Input}{Input}\SetKwInOut{Output}{Output}

  \Input{Datasets from $K$ clients: $\mathcal{D}^{1}, \mathcal{D}^{2},\cdots,\mathcal{D}^{K}= \{ (\mathbf{b}_{i}, \mathbf{m}_{i})\mid i=1,\cdots ,N_{K} \}  $; number of communication epochs $T$; number of local epochs $Z$; learning rate for client $k$: $\eta_{k}$; initialize $\mathcal{L}_{C}^{k}=0$ and $\alpha_{C^{k}}^{t}=\frac{N_{k}}{N}$\;}
  \Output{the final server parameters: $\Theta_{C}^{T}$; client parameters: $\Theta_{C_{g}}^{T}\cup \Theta_{C_{l}^{k}}^{T,Z}$\;}
  \BlankLine
  \For{$t=1,2,\cdots, T$}{
    \For{$k=1,2,\cdots, K$}{
      \label{forins}
      receive the global weights $\Theta_{C}^{t}$ from the server\;
      use $\Theta_{C}^{t}$ to calculate $\mathcal{L}_{C}^{k}$ on the training subset 2 of client $k$ based on Eq. (11)\;
      use $\Theta_{C_{g}}^{t}\cup \Theta_{C_{l}^{k}}^{t}$ to calculate $\mathcal{L}_{C_{s2}^{k}}$ based on Eq. (15)\;
      \For{$z=1,2,\cdots, Z$}{
      calculate $\bigtriangledown \mathcal{L}_{C_{rec}^{k}}$ based on Eq. (10)\;
      $  \Theta_{C^{k}}^{t,z+1}= \Theta_{C^{k}}^{t,z}-\eta_{k}\bigtriangledown\mathcal{L}_{C_{rec}^{k}} $\;
      }
      upload weights $\Theta_{C^{k}}^{t}$ and $\mathcal{L}_{C_{s2}^{k}}$to server\;
    }
    receive $\Theta_{C^{k}}^{t}$ and $\mathcal{L}_{C_{s2}^{k}}$ from client\;
    calculate $\alpha_{C^{k}}^{t}$ based on Eq. (14)\;
    aggregate $\Theta_{C}^{t}=\sum_{k=1}^{K}\alpha _{C^{k}}^{t}\cdot \Theta _{C^{k}}^{t} $\;
    distribute $\Theta_{C}^{t}$ to each client\;
  }
\end{algorithm}

\subsection{Spatial Laplacian Attention Module}
The residual spatial Laplacian attention module (RSLAM) is shown in Fig. 2(b), which consists of four continuous conv-conv-relu, a spatial Laplacian attention module (SLAM), a conv, and jump connection. It is mainly used to remove the noise from the image, i.e., $E_{\bm{\omega}}(\cdot)$ in Eq. (3). The current MR image reconstruction methods are not accurate enough and details are easily lost. The importance of information in different channels and locations in the neural networks is different, and the learning of important features can be facilitated by introducing an attention mechanism. To capture the detailed information of the image, we introduce a spatial Laplacian attention module, whose detailed structure is shown in Fig. 2. SLAM consists of a spatial attention module and a Laplacian attention module. The details of spatial attention module are as follows:

\begin{equation}
    \mathcal{M}_{s}(\mathbf{F})=\sigma (f_{7\times 7}(cat(AvgPool(\mathbf{F}),Maxpool(\mathbf{F})))) 
\end{equation}
where $\sigma$ denotes $sigmoid$ function, $\mathbf{F}\in \mathbb{R}^{C\times H\times W} $ denotes the feature map, and $cat(\cdot)$ denotes the concatenate operation. The spatial attention module focuses on the location of important features.

a) \textbf{Laplacian Attention Module:} 
Laplacian attention (Fig. 2(a)) mainly focuses on which channel is important, and effectively increases the receptive field of the network by introducing dilated convolutions. First, the global descriptor $\mathbf{g}_{avg}\in \mathbb{R}^{C\times 1\times 1} $ and $\mathbf{g}_{max}\in \mathbb{R}^{C\times 1\times 1} $ are obtained by adaptive average pooling and adaptive maximum pooling. Key features at different scales are then learned by the Laplacian pyramid with dilated convolutions:

\begin{equation} 
 \begin{split}
\mathbf{g}_{avg}^{p}(\mathbf{F})=cat(\tau (D_{f_{3}}(\mathbf{g}_{avg}(\mathbf{F}))),\tau (D_{f_{5}}(\mathbf{g}_{avg}(\mathbf{F}))),\\
\tau (D_{f_{7}}(\mathbf{g}_{avg}(\mathbf{F}))))
\end{split} 
\end{equation}

\begin{equation}
 \begin{split}
    \mathbf{g}_{max}^{p}(\mathbf{F})=cat(\tau (D_{f_{3}}(\mathbf{g}_{max}(\mathbf{F}))),\tau (D_{f_{5}}(\mathbf{g}_{max}(\mathbf{F}))), \\
    \tau (D_{f_{7}}(\mathbf{g}_{max}(\mathbf{F}))))
    \end{split} 
\end{equation}
where $\tau$ denotes the $ReLU$ operation, $D_{f_{3}}(\cdot ), D_{f_{5}}(\cdot ), D_{f_{7}}(\cdot )$ denote the dilated convolutions with scales 3, 5, 7, respectively. To adjust the channel number of the feature map, one convolution is utilized for $\mathbf{g}_{avg}^{p}$ and $\mathbf{g}_{max}^{p}$ respectively, and then the convolved results are summed and fed into the $sigmoid$ function to obtain the statistical information of different channels.

\begin{equation}
    \mathcal{M}_{L}(\mathbf{F})=\sigma (f_{3\times 3}(\mathbf{g}_{avg}^{p}(\mathbf{F}))+f_{3\times 3}(\mathbf{g}_{max}^{p}(\mathbf{F})))
\end{equation}
Ultimately, the spatial Laplacian attention is calculated as follows:

\begin{equation}
    \mathcal{M}_{SL}(\mathbf{F})=\mathcal{M_{S}}(\{\mathcal{M}_{L}(\mathbf{F})\otimes \mathbf{F} \})\otimes  \{\mathcal{M}_{L}(\mathbf{F})\otimes \mathbf{F} \}
\end{equation}
where $\otimes$ denotes element-wise multiplication.

\begin{figure}[htbp]
    \centering
    \setlength{\abovecaptionskip}{0.cm}
    \includegraphics[width=8.5cm, keepaspectratio]{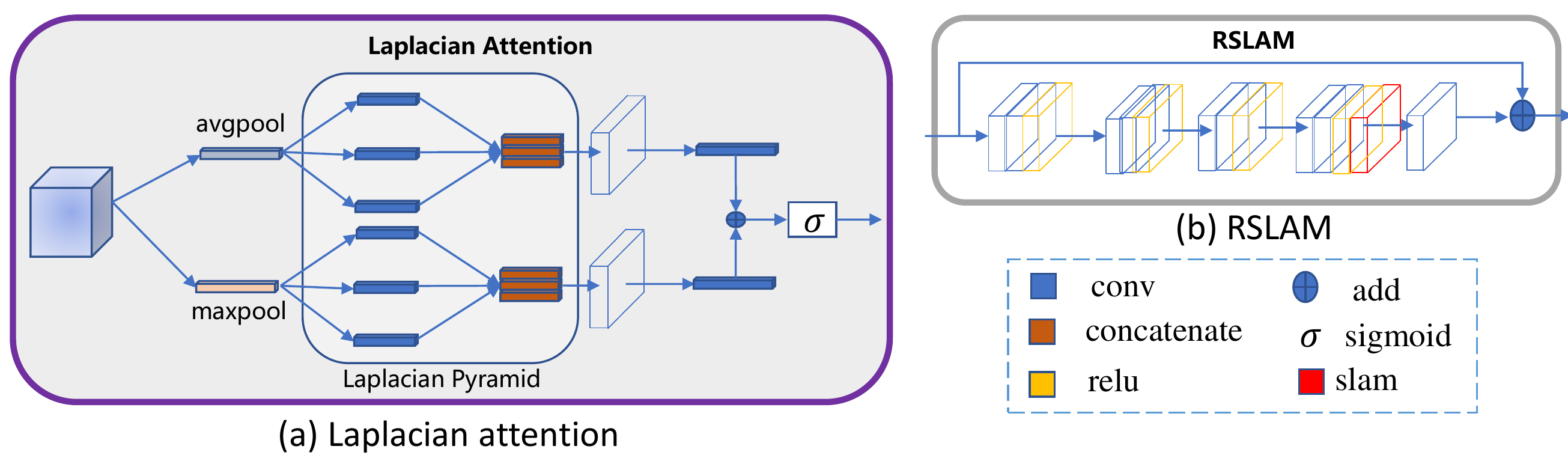}
    \caption{Attention module. (a) Laplacian attention. (b) Residual spatial Laplacian attention module (RSLAM).
    }
    \label{fig2}
\end{figure}

\section{Experimental Results and Analysis}
In this section, we will introduce the datasets, implementation details, comparison methods, evaluation metrics, and experimental results. Our experiments mainly include: \textbf{1}) performance comparison (\textbf{Scenario 1}: Comparing the reconstruction performance of different state-of-the-art methods at $4\times$ acceleration rate with 1D random sampling; \textbf{Scenario 2}: Testing the performance of different methods under greater data heterogeneity, including 1D uniform sampling, 1D random sampling and 2D random sampling at $4\times$ acceleration rate) and generalization analysis; \textbf{2}) The experiment about noise under \textbf{Scenario 1} to compare the robustness of different methods; \textbf{3}) Ablation study.

\subsection{Experimental Settings}
\subsubsection{Datasets}
One in-house dataset \cite{wang2020deepcomplexmri} (was obtained with a 3D TSE protocol by the United Imaging system, uMR 790) and two public datasets (fastMRI \cite{knoll2020fastmri}, CC359\footnote{https://sites.google.com/view/calgary-campinas-dataset/mr-reconstruction-challenge}) were utilized in this study to evaluate the effectiveness of the proposed method. Each dataset is divided into training subset 1, subset 2 and testing set. Details of the datasets are provided as follows: 
1) \textbf{In-house dataset \cite{wang2020deepcomplexmri}:} T1-, T2-, and PD-weighted MRI scans of 22 subjects were acquired using a United Imaging system, uMR 790. T1-weighted images were acquired using the following scan parameters: TE/TR = 11/928 ms, voxel resolution = 0.9×0.9×0.9 mm. Similarly, for T2-weighted images, the scan parameters: TE/TR = 149/2500 ms, voxel resolution = 0.9×0.9×0.9 mm. For PD-weighted images, the scan parameters: TE/TR = 13/2000 ms, voxel resolution = 1.0×1.1×1.1 mm. All data are cropped to a matrix size of 256×256. Informed consent was obtained from every subject in compliance with the Institutional Review Board policy \cite{wang2020deepcomplexmri}; 2) \textbf{fastMRI \cite{knoll2020fastmri}:} T1-weighted images of 3443 subjects are utilized; 3) \textbf{CC359:}
 This dataset provides 35 fully-sampled T1-weighted MR scans (6224 slices) acquired on a clinical MR scanner (Discovery MR 750; General Electric (GE) Healthcare, Waukesha, WI). The matrix size is 256×256.  

\subsubsection{Implementation details}
The server and client models adopt the same network architectures, which consist of alternating RSLAM and DC blocks (Fig. 1). In our implementation, there are 5 alternating modules. To deal with the complex MRI data, we superimpose the real and imaginary parts by channel. All the networks were trained using the PyTorch framework with two NVIDIA RTX A6000 GPUs (each with 48 GB memory). In training phase, we employed the AdamW optimizer. Networks were trained for 220 global epochs with 2 local epochs. The initial learning rate was set to $1\times10^{-4}$, and the batch size was 24. $\gamma_{k}$ in Eq. (10) was set to 0.1.

\subsubsection{Comparision Methods}
We compared the results of ModFed with those of various state-of-the-art federated learning methods to demonstrate the effectiveness of our method. The comparison methods include FedAvg \cite{mcmahan2017communication}, FedProx \cite{li2020fedprox}, LG-FedAvg \cite{liang2020think}, FedPer \cite{arivazhagan2019federated}, and FedMRI \cite{feng2022specificity}.  We also include one non-federated learning comparison method. The details of different comparison methods can be found in the published papers. Here, we introduce the essence of each method for a quick grasp:
1) \textbf{SingleSet:} each client is trained using their local data without using federated learning; 2) \textbf{FedAvg:} train a global model by averaging the weights from all local clients; 3) \textbf{FedProx:} train a global model by adding a proximal term to the objective function of the local clients; 4) \textbf{LG-FedAvg:} learn different local heads and one global-shared network body; 5) \textbf{FedPer:}  personalized federated learning method that introduces a personalized layer after the base layer; 6) \textbf{FedMRI:} personalized federated learning method that introduces a weighted contrastive regularization to correct any deviation between the client and server for all clients during optimization.

\subsubsection{Evaluation Metrics}
Two metrics, peak signal-to-noise ratio (PSNR) and structural similarity (SSIM) \cite{wang2004image}, were calculated to quantitatively evaluate the experimental results. The SSIM index is the product of the luminance, contrast and structure measure functions. The formula is:
\begin{equation}
    SSIM(\mathbf{X},\mathbf{Y})=\frac{(2\mu_{x}\mu_{y}+C_{1})(2\delta_{xy}+C_{2})}
{(\mu_{x}^{2}+\mu_{y}^{2}+C_{1})(\delta_{x}^{2}+\delta_{y}^{2}+C_{2})} 
\end{equation}
where $\mu_{x}$ and $\mu_{y}$ are the average of $\mathbf{X}$ and $\mathbf{Y}$, respectively; $\delta_{x}^{2}$ and $\delta_{y}^{2}$ are the variance of $\mathbf{X}$ and $\mathbf{Y}$, respectively; $\delta_{xy}$ are the covariance of $\mathbf{X}$ and $\mathbf{Y}$; $C_{1}=(k_{1}L)^{2}$ and $C_{2}=(k_{2}L)^{2}$ are constants used to ensure stability; $L$ is the dynamic range of pixel values.

\subsection{Experimental Results}
\subsubsection{Comparison to State-of-the-Art Federated Learning Methods}
To comprehensively demonstrate the effectiveness of the proposed ModFed, we compare it with the state-of-the-art FL methods under two scenarios. \textbf{Scenario 1:} Comparing the reconstruction performance of different methods at 4$\times$ acceleration rate with 1D random sampling. \textbf{Scenario 2:} Testing the performance of different methods under greater data heterogeneity. Under this scenario, different sampling patterns are utilized for the three datasets to simulate the condition when different scanners are used. Specifically, 1D uniform sampling with 4$\times$ acceleration is used for fastMRI. 1D random sampling with 4$\times$ acceleration is used for CC359. And 2D random sampling with 4$\times$ acceleration is used for the in-house dataset.

a) \textbf{Results under Scenario 1}: Figs. 3 and 4 show the quantitative results of all methods under \textbf{Scenario 1}. Overall, ModFed achieves competitive performance when compared to the other methods on three datasets. The classical federated learning method, FedAvg, gives common results. The four personalized federated learning methods (LG-FedAvg, FedPer, FedMRI, and our ModFed) obtain better reconstruction performance than classical federated learning approaches (FedAvg and FedProx), validating their effectiveness in addressing the data heterogeneity issue. These methods have successfully learned a better model for each client. Nevertheless, our proposed ModFed can still generate higher PSNR and SSIM results than LG-FedAvg, FedPer, and FedMRI on three datasets. This observation confirms that ModFed can effectively solve the data heterogeneity issue. Moreover, ModFed generates better results than SingleSet, showing its ability to handle data with large variations in data distribution while preventing data privacy leakage.

\begin{figure}[htbp]
    \centering
    \setlength{\abovecaptionskip}{0.cm}
    \includegraphics[width=8.5cm, keepaspectratio]{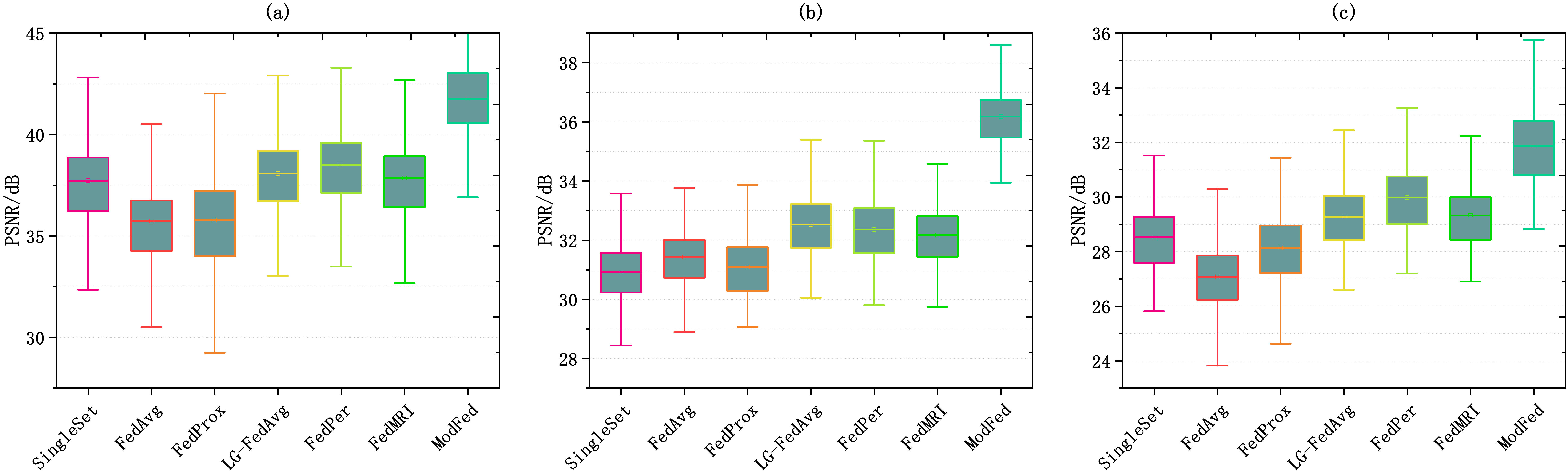}
    \caption{PSNR of reconstructed MR images of different methods on the three datasets ((a) fastMRI, (b) CC359, (c) in-house dataset) under \textbf{Scenario 1}. The centerline and error bar of the box plots represent the median and interquartile range (25th-75th percentile).    }
    \label{fig3}
\end{figure}

\begin{figure}[htbp]
    \centering
    \setlength{\abovecaptionskip}{0.cm}
    \includegraphics[width=8.5cm, keepaspectratio]{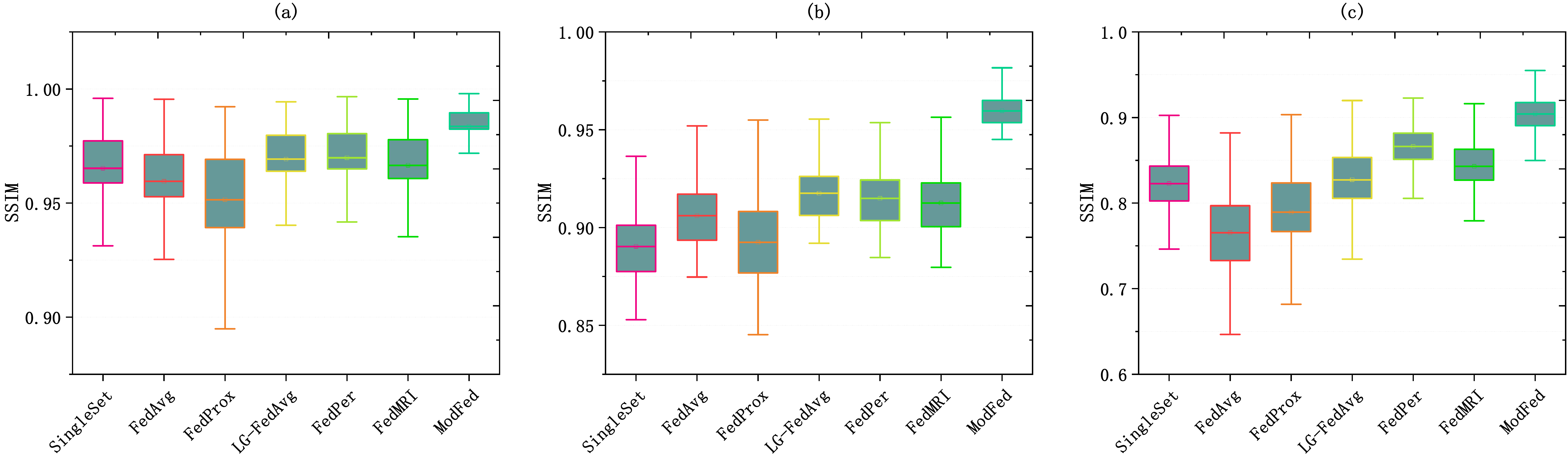}
    \caption{SSIM of reconstructed MR images of different methods on the three datasets ((a) fastMRI, (b) CC359, (c) in-house dataset) under \textbf{Scenario 1}. The centerline and error bar of the box plots represent the median and interquartile range (25th-75th percentile).
    }
    \label{fig4}
\end{figure}

\begin{figure*}[htbp]
    \centering
    \setlength{\abovecaptionskip}{0.cm}
    \includegraphics[width=20cm, height=10cm, keepaspectratio]{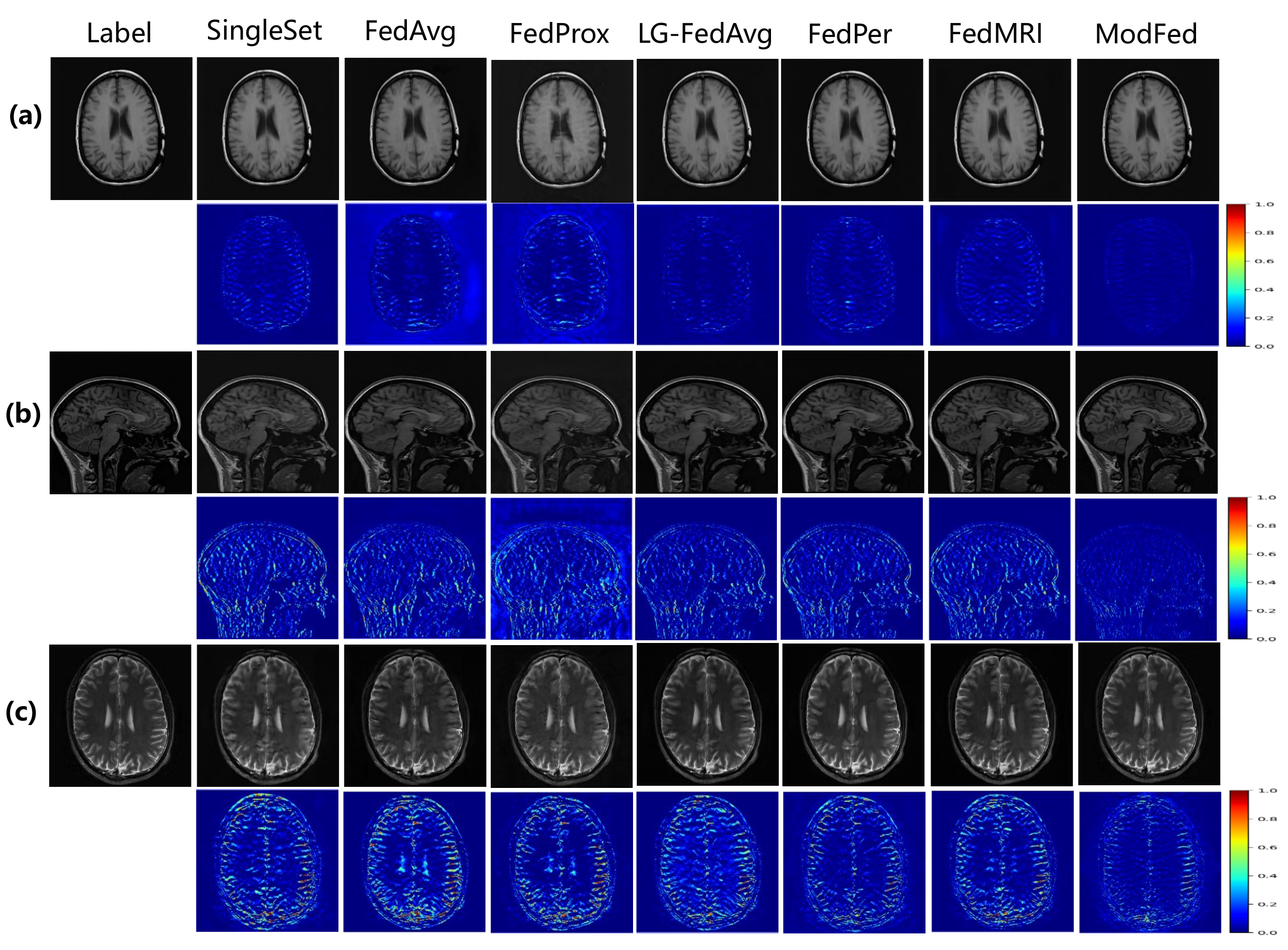}
    \caption{Qualitative reconstruction results of different methods on the three datasets ((a) fastMRI, (b) CC359, and (c) in-house dataset) under \textbf{Scenario 1}. From left to right, the eight images correspond to the reference image, and the reconstructed images of Singleset, FedAvg, FedProx, LG-FedAvg, FedPer, FedMRI, and our ModFed, respectively. The second, fourth and sixth rows plot the corresponding error maps.
    }
    \label{fig5}
\end{figure*}

In addition to the quantitative statistics, example reconstructed MR images as well as the corresponding error maps are plotted to qualitatively evaluate the reconstruction performance of different methods (Fig. 5). Overall, similar conclusions can be made as those of the quantitative results that ModFed can reconstruct MR images with a higher quality and smaller errors. According to the plotted error maps, ModFed can achieve better reconstruction results than FedMRI. One possible limitation of FedMRI is that it still uses the average aggregation method, which treats all clients fairly without differentiating the contributions of each client to the model training. In contrast, ModFed effectively mitigates this problem by introducing an adaptive dynamic aggregation scheme. As a result, more weight is enforced on the clients with larger losses, and the problem of date heterogeneity can be properly addressed. The introduced adaptive dynamic aggregation scheme guarantees that the model can maintain good performance even on datasets with data heterogeneity. Moreover, personalized client-loss regularization further enhances the reconstruction performance of each client. Therefore, both quantitative and qualitative results suggest that our proposed ModFed can properly handle the data heterogeneity issue under \textbf{Scenario 1}.

\begin{figure}[htbp]
    \centering
    \setlength{\abovecaptionskip}{0.cm}
    \includegraphics[width=8.5cm, keepaspectratio]{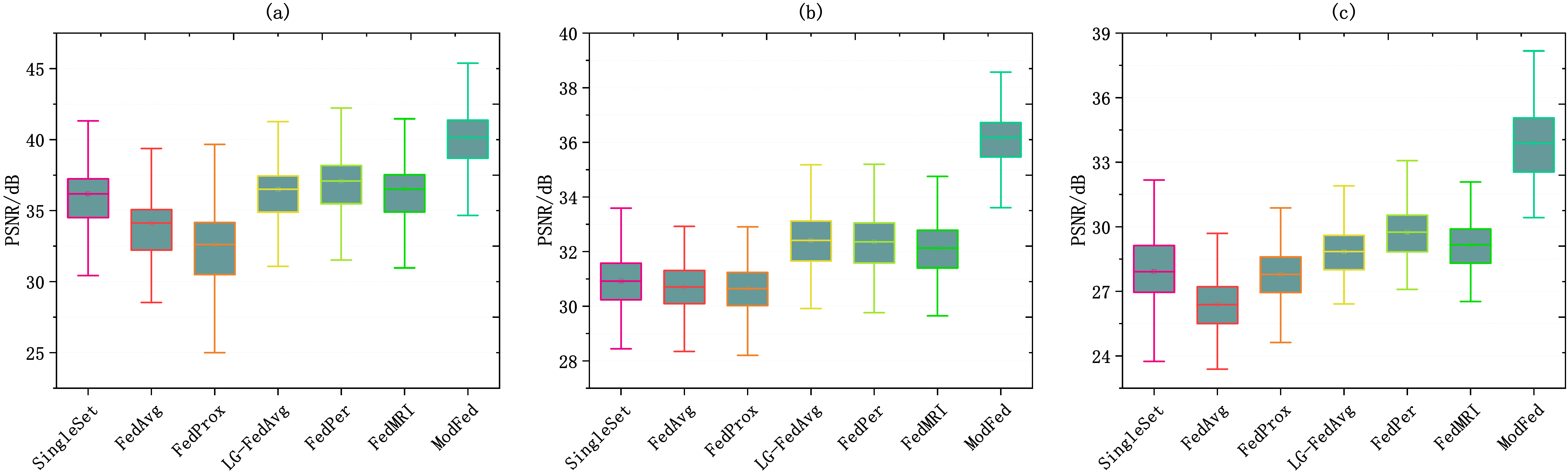}
    \caption{PSNR of reconstructed MR images of different methods on the three datasets ((a) fastMRI, (b) CC359, and (c) in-house dataset) under \textbf{Scenario 2}. The centerline and error bar of the box plots represent the median and interquartile range (25th-75th percentile).
    }
    \label{fig6}
\end{figure}

\begin{figure}[htbp]
    \centering
    \setlength{\abovecaptionskip}{0.cm}
    \includegraphics[width=8.5cm, keepaspectratio]{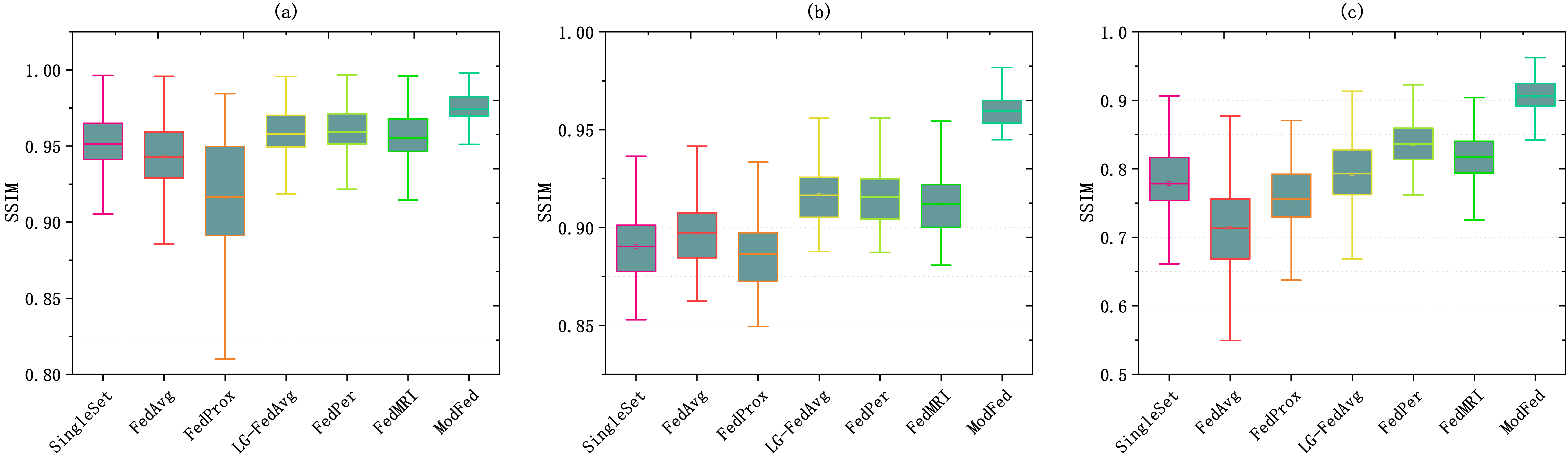}
    \caption{SSIM of reconstructed images of different methods on the three datasets ((a) fastMRI, (b) CC359, and (c) in-house dataset) under \textbf{Scenario 2}. The centerline and error bar of the box plots represent the median and interquartile range (25th-75th percentile).
    }
    \label{fig7}
\end{figure}

\begin{figure*}[htbp]
    \centering
    \setlength{\abovecaptionskip}{0.cm}
    \includegraphics[width=20cm,height=10cm,keepaspectratio]{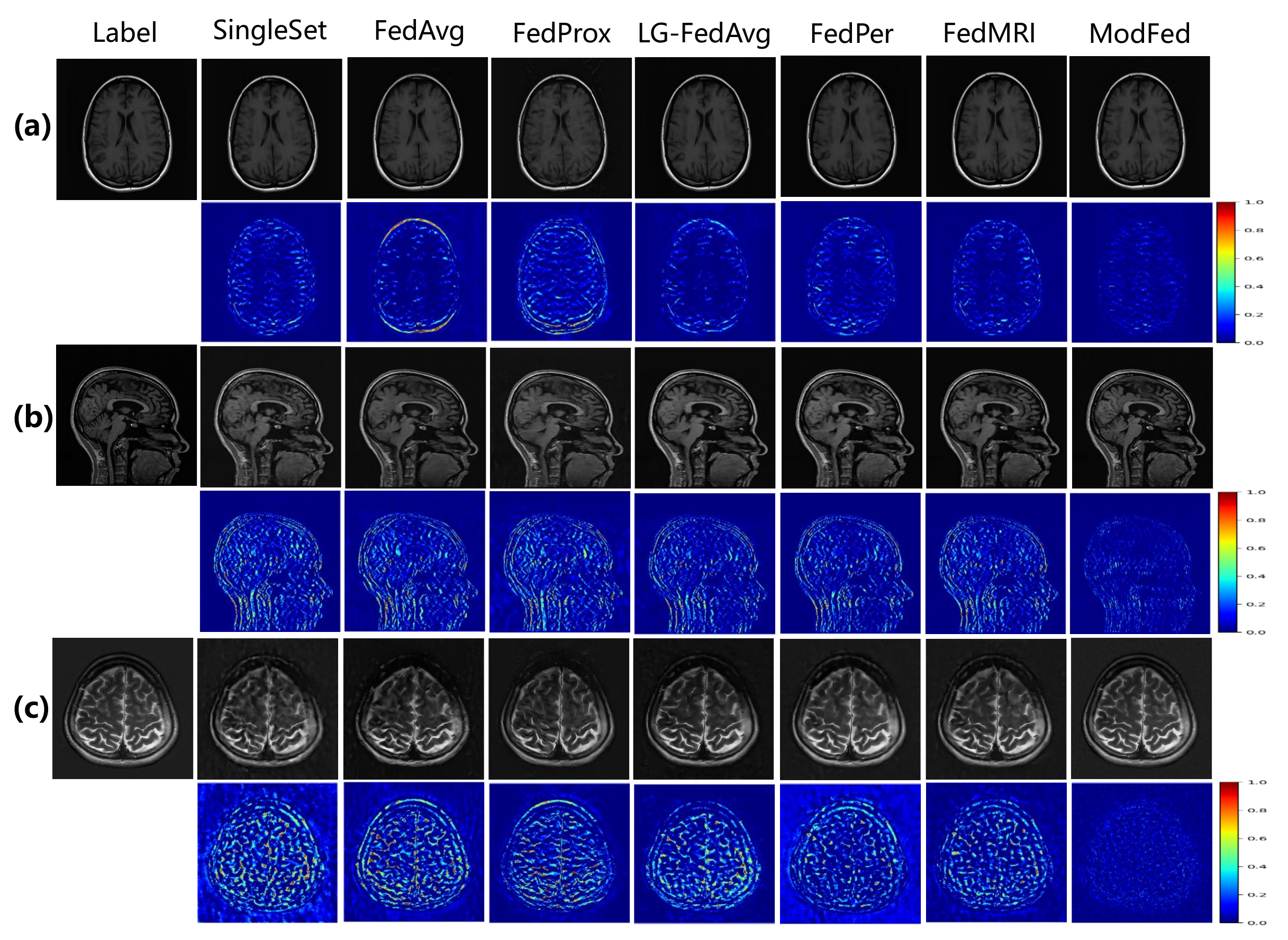}
    \caption{Qualitative reconstruction results of different methods on the three datasets ((a) fastMRI, (b) CC359, and (c) in-house dataset) under \textbf{Scenario 2}. Results are shown for reference image, Singleset, FedAvg, FedProx, LG-FedAvg, FedPer, FedMRI, along with our method ModFed. The second, fourth and sixth rows indicate the corresponding error maps.
    }
    \label{fig8}
\end{figure*}

b) \textbf{Results under Scenario 2}: To comprehensively evaluate the performance of the proposed method, we conducted experiments under \textbf{Scenario 2}, which represents a situation with greater data heterogeneity. Figs. 6 and 7 give the quantitative results of the different methods on the three datasets. Statistics show that FedProx has worse reconstruction performance, which could be caused by the increased data heterogeneity. Personalized federated learning methods (LG-FedAvg, FedPer, FedMRI, and our ModFed) achieve relatively better results. Our proposed ModFed generates the highest PSNR and SSIM results under this scenario. By adaptive dynamic aggregation scheme and personalized client-side loss regularization, ModFed can effectively conquer the greater data heterogeneity issue.

Example reconstructed MR images as well as the corresponding error maps of different methods are plotted in Fig. 8. Similarly, we can observe that ModFed can reconstruct MR images with a higher quality and smaller errors. Both quantitative and qualitative results indicate that ModFed is able to address the data heterogeneity issue under both scenarios and reconstruct MR images with high quality. The experimental results show ModFed is a highly effective personalized federated learning method that can handle different scales of data heterogeneity issue.

\subsubsection{Generalization Capability of ModFed}
A standard generalization bound \cite{nagarajan2021explaining} of a model $f_{\Theta}$ can be expressed as:

\begin{equation}
   \mathfrak{L}_{D}(f_{\Theta}) \le \mathfrak{L}_{S}(f_{\Theta})+ O (\sqrt{\frac{p}{m }} )
\end{equation}
where $ \mathfrak{L}_{D}(f_{\Theta})$ denotes the error on the testing set, $\mathfrak{L}_{S}(f_{\Theta})$ denotes the error on the training set, $p$ is the number of parameters of the model, $m$ is the number of training samples. Since the training data $m$ is fixed, the testing error of the model is only related to the training error and the number of model parameters $p$. To be specific, the testing error is proportional to the training error and the number of model parameters. Figs. 9 and 10 plot the comparison results of the training error and the number of model parameters of different methods on the three datasets under \textbf{Scenario 1} and \textbf{Scenario 2}, respectively. According to the experimental results, our method obtains the minimum training error and the number of model parameters under both scenarios and obtains the best results on the in-house dataset under \textbf{Scenario 2}. Since the in-house dataset consists of data acquired with different sequences, it is more heterogeneous than the other two datasets. Our proposed ModFed can maintain its highly competitive generalization capability on all three datasets.

\begin{figure}[htbp]
    \centering
\setlength{\abovecaptionskip}{0.cm}
\includegraphics[width=8.5cm,keepaspectratio]{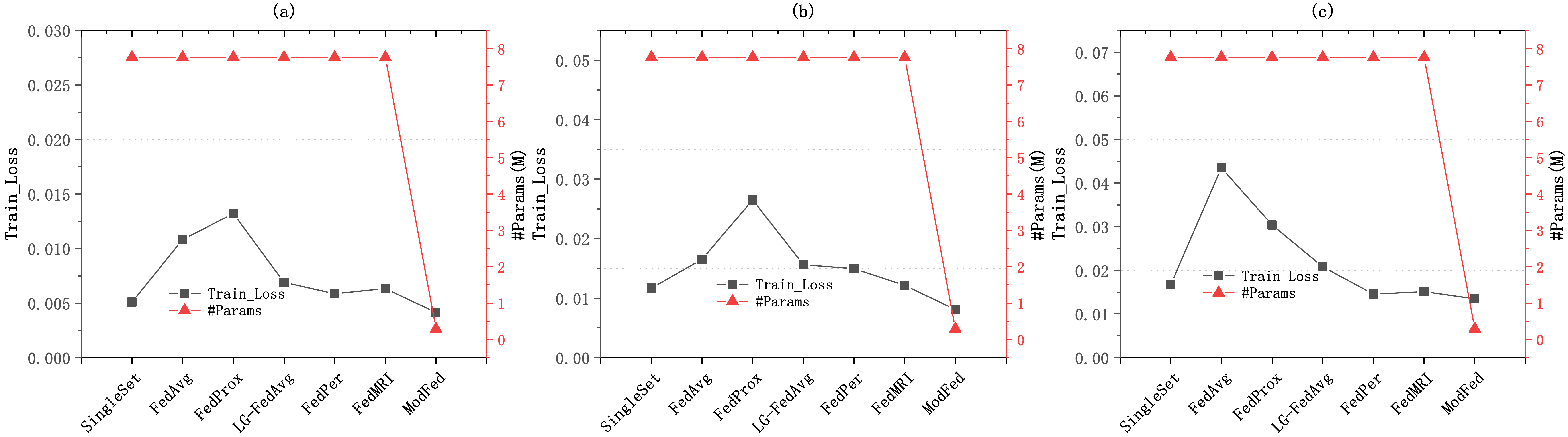}
    \caption{Training error and model parameter (in millions) quantity of different methods on the three datasets ((a) fastMRI, (b) CC359, and (c) in-house dataset) under \textbf{Scenario 1}.
    }
    \label{fig9}
\end{figure}

\begin{figure}[htbp]
    \centering
\setlength{\abovecaptionskip}{0.cm}
\includegraphics[width=8.5cm, keepaspectratio]{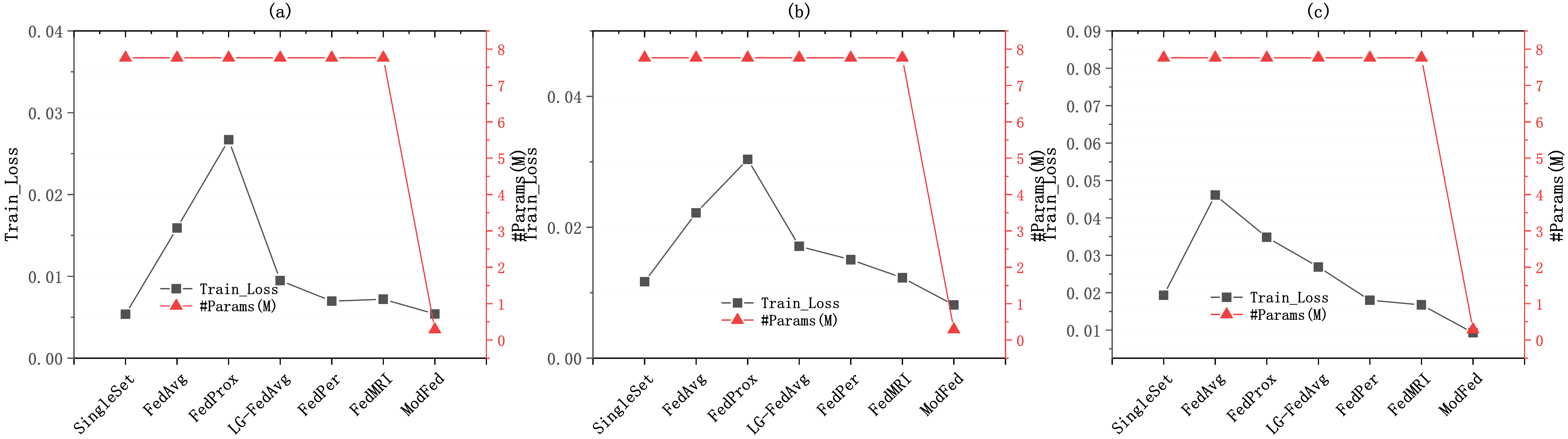}
    \caption{Training error and model parameter (in millions) quantity of different methods on the three datasets ((a) fastMRI, (b) CC359, and (c) in-house dataset) under \textbf{Scenario 2}.
    }
    \label{fig10}
\end{figure}

\subsubsection{Robustness to Noise}
More experiments were conducted to investigate the robustness of different methods when complex white Gaussian noise is added to the k-space data. Figs. 11 and 12 give the PSNR and SSIM values of the different methods on the three datasets. From the results, we can see the proposed ModFed can still maintain its encouraging performance when noise is added and have the best PSNR and SSIM values on the three datasets. This set of experiments validates that ModFed is robust to noise when compared to the comparison methods.

\begin{figure}[htbp]
    \centering
    \setlength{\abovecaptionskip}{0.cm}
\includegraphics[width=8.5cm,keepaspectratio]{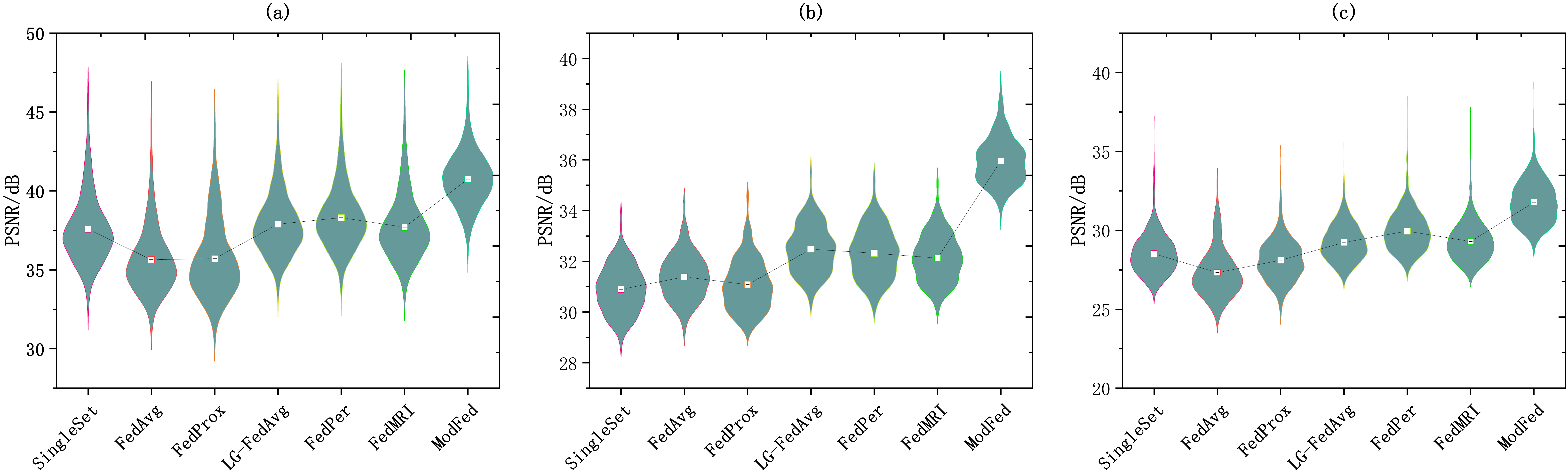}
    \caption{PSNR of reconstructed MR images of different methods on the three datasets ((a) fastMRI, (b) CC359, and (c) in-house dataset) under \textbf{Scenario 1} when white Gaussian noise (with a variance of 0.03) is added to the k-space data points. Violin plots indicate the distribution states and probability densities. The mean PSNR of each method is connected by the black line for easy comparison.
    }
    \label{fig11}
\end{figure}

\vspace{1cm}
\begin{figure}[htbp]
    \centering
    \setlength{\abovecaptionskip}{0.cm}
\includegraphics[width=8.5cm,keepaspectratio]{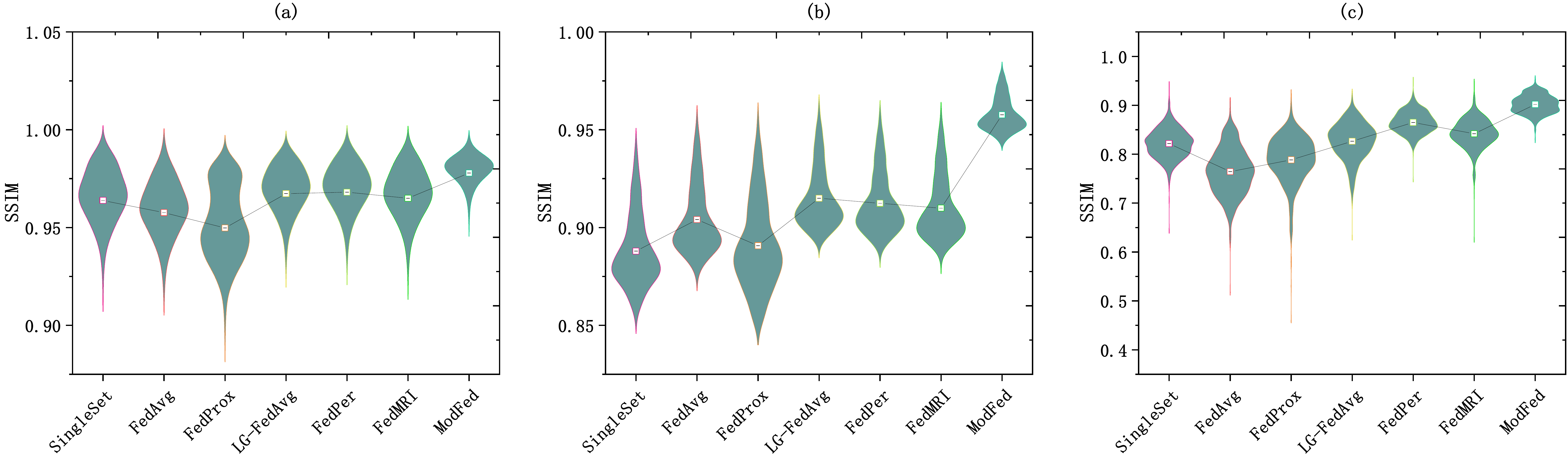}
    \caption{SSIM of reconstructed MR images of different methods on the three datasets ((a) fastMRI, (b) CC359, and (c) in-house dataset) under \textbf{Scenario 1} when white Gaussian noise (with a variance of 0.03) is added to the k-space data points. Violin plots indicate the distribution states and probability densities. The mean SSIM of each method is connected by the black line for easy comparison.
    }
    \label{fig12}
\end{figure}

\begin{figure}[htbp]
    \centering
    \setlength{\abovecaptionskip}{0.cm}
\includegraphics[width=8cm, keepaspectratio]{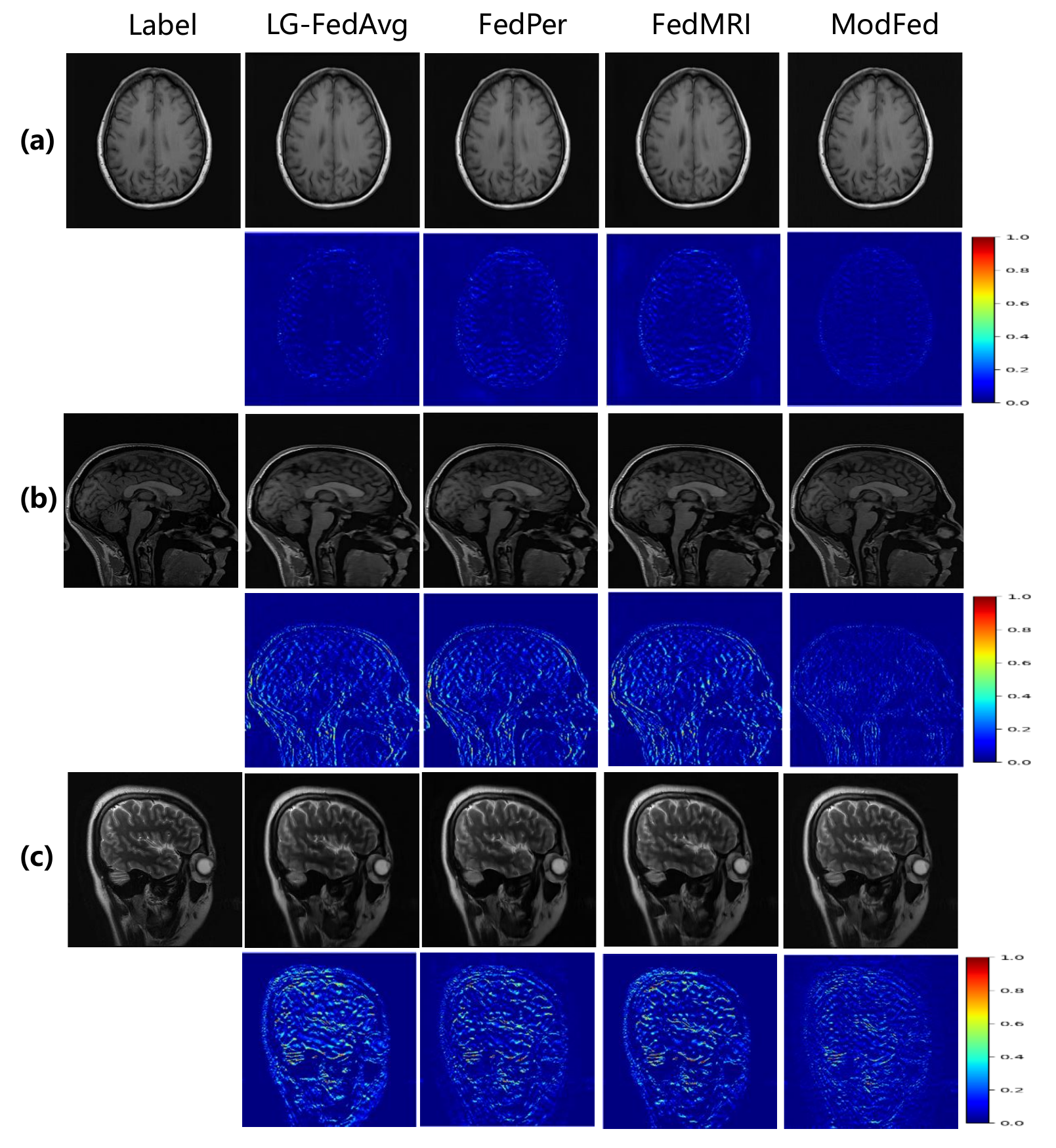}
    \caption{Qualitative reconstruction results of different methods on the three datasets ((a) fastMRI, (b) CC359, and (c) in-house dataset) under \textbf{Scenario 1} when white Gaussian noise (with a variance of 0.03) is added to the k-space data points. From left to right, the five images correspond to the reference image, and the reconstructed images of LG-FedAvg, FedPer, FedMRI, and our ModFed. The second, fourth, and sixth rows show the corresponding error maps.}
    \label{fig13}
\end{figure}

In addition, we compare the results achieved by ModFed to those generated by the other three personalized methods (LG-FedAvg, FedPer, and FedMRI) qualitatively under \textbf{Scenario 1} when white Gaussian noise (with a variance of 0.03) is added (Fig. 13). According to the plotted error maps, ModFed has better suppression ability to noise than the other three methods. 

\subsubsection{Ablation Study}
In this section, we analyze the contributions of the different components of ModFed to enhance reconstruction performance. One baseline model and three variants of ModFed were implemented under \textbf{Scenario 1}. The baseline model is trained using the classical federated average aggregation method. M1 contains only the spatial Laplacian attention module. M2 contains only the personalized client-side loss regularization. M3 contains only the adaptive dynamic aggregation scheme. Table 1 lists the results of different methods. Overall, it can be observed that all three variants can achieve enhanced reconstruction performance when compared to the baseline, showing the positive effects of the three components (the spatial Laplacian attention module, the personalized client-side loss regularization, and the adaptive dynamic aggregation scheme). By enabling all three components, our final model, ModFed, achieves the best results. Experimental results show that all three components contribute to the reconstruction results.

\begin{table*}[htbp]
\centering
    \caption{Results of ablation studies under \textbf{Scenario 1}}
    \label{tab1}
    \resizebox{\textwidth}{!}{
\tiny 
\begin{tabular}{cccccccccccc}
\toprule[0.5pt]
\multirow{2}{*}{Methods} & \multirow{2}{*}{attention} & \multirow{2}{*}{$\mathcal{L}_{C}^{k}$} & \multirow{2}{*}{$\alpha_{C^{k}}^{t}$} & \multicolumn{2}{c}{fastMRI} & \multicolumn{2}{c}{CC359} & \multicolumn{2}{c}{in-house} & \multicolumn{2}{c}{Average} \\
\cmidrule(l){5-12} 
\multicolumn{1}{c}{}  &  &   &   & 
PSNR$\uparrow$  & 
SSIM$\uparrow$  & 
PSNR$\uparrow$  & 
SSIM$\uparrow$  & 
PSNR$\uparrow$  & 
SSIM$\uparrow$  & 
PSNR$\uparrow$  & 
SSIM$\uparrow$    \\ 
\midrule[0.5pt]
baseline & \usym{2718} & \usym{2718} & \usym{2718} & 41.5936 & 0.9834 & 35.9904 & 0.9578 & 31.7090  & 0.9022  & 36.4310 & 0.9478      
\\
M1 & \usym{2714}  & \usym{2718}  & \usym{2718}  & 41.7222  & 0.9835  & 36.0752  & 0.9587  & 31.7356  & 0.9023  & 36.5110   & 0.9482      
\\
M2  & \usym{2718}  & \usym{2714}  & \usym{2718}  & 41.6344   & 0.9835   & 36.1267  & 0.9593     & 31.8228  & 0.9041  & 36.5280  & 0.9490       
\\
M3  & \usym{2718}  & \usym{2718}  & \usym{2714}  & 41.6471  & 0.9835  & 36.1786  & 0.9597  & 31.8182  & 0.9039  & 36.5480  & 0.9490       
\\
\textbf{ModFed}  & \usym{2714}  & \usym{2714}  & \usym{2714}  & \textbf{41.7607}   & \textbf{0.9836}  & \textbf{36.1826}  & \textbf{0.9597}     & \textbf{31.8600} & \textbf{0.9042} & \textbf{36.6011} & \textbf{0.9492}
\\
\bottomrule[0.5pt]
\end{tabular}
}
\end{table*}

\section{Conclusion}
In this work, we investigated federated learning method development to achieve the joint learning of multi-institutional data where data heterogeneity exists among the clients. To this end, we propose a model-driven federated learning framework for accurate MR image reconstruction from undersampled k-space data (ModFed), which relieves each client’s dependency on large data. Specifically, an adaptive dynamic aggregation scheme is used to address the data heterogeneity issue and improve the generalization capability and robustness of the trained model. A spatial Laplacian attention mechanism and a personalized client-side loss regularization are used to capture the detailed information of the image, achieving enhanced reconstruction performance on each client. Experimental results validated that ModFed can achieve better MR image reconstruction on the three datasets when compared to five state-of-the-art federated learning methods.

\nocite{*}
\bibliography{/main} 
\bibliographystyle{IEEEtran} 

\end{document}